\documentclass[a4paper,fleqn]{cas-sc}

\usepackage[numbers]{natbib}
\usepackage{subcaption}
\usepackage{amsmath}
\usepackage{url}
\usepackage{lineno}
\usepackage{nccmath}

\def\tsc#1{\csdef{#1}{\textsc{\lowercase{#1}}\xspace}}
\tsc{WGM}
\tsc{QE}
\tsc{EP}
\tsc{PMS}
\tsc{BEC}
\tsc{DE}
\usepackage{enumerate}
\begin{document}
\let\WriteBookmarks\relax
\def\floatpagepagefraction{1}
\def\textpagefraction{.001}
\shorttitle{Stochastic modeling of a neutron imaging center at the Brazilian Multipurpose Reactor}
\shortauthors{LP de Oliveira, APS Souza, FA Genezini and A dos Santos}

\title [mode = title]{Stochastic modeling of a neutron imaging center at the Brazilian Multipurpose Reactor}                      
\author[1]{LP de Oliveira}
\cormark[1]
\ead{oliveira.phys@gmail.com}

\author[2]{APS Souza}

\author[1]{FA Genezini}

\author[1]{A dos Santos}

\address[1]{Instituto de Pesquisas Energéticas e Nucleares, IPEN/CNEN, Av. Prof. Lineu Prestes, 2242 – Cidade Universitária – CEP 05508-000 São Paulo – SP – Brazil}

\address[2]{Idaho National Laboratory, INL, 1955 N. Fremont Ave. Idaho Falls, ID 83415, United States of America}

\cortext[cor1]{Corresponding author}
\begin{abstract}
Neutron imaging is a non-destructive technique for analyzing a wide class of samples, such as archaeological or industrial material structures. In recent decades, technological advances have had a great impact on the neutron imaging technique, which has meant an evolution from simple radiographs using films (2D) to modern tomography systems with digital processing (3D). The 5 MW research nuclear reactor IEA-R1, which is located at the Instituto de Pesquisas Energéticas e Nucleares (IPEN) in Brazil, possesses a neutron imaging instrument with $1.0 \times 10^{6}$ $n/cm^{2}s$ in the sample position. IEA-R1 is over 60 years old and the future of neutron science in Brazil, including imaging, will be expanded to a new facility called the Brazilian Multipurpose Reactor (RMB, Portuguese acronym), which will be built soon. The new reactor will house a suite of instruments at the Neutron National Laboratory, including the neutron imaging facility, viz., Neinei. Inspired by recent author's works, we model the Neinei instrument through stochastic Monte Carlo simulations. We investigate the sensitivity of the neutron imaging technique parameter ($L/D$ ratio) with the neutron flux, and the results are compared to data from the Neutra (PSI), Antares (FRM II), BT2 (NIST) and DINGO (OPAL) instruments. The results are promising and provide avenues for future improvements.
\end{abstract}

\begin{keywords}
Neutron Imaging \sep Monte Carlo \sep MCSTAS \sep
\end{keywords}

\maketitle
\section{Introduction}
 
The Brazilian Multipurpose Reactor (RMB) project inaugurates a new era of research on nuclear science and technologies in Brazil. Besides radiopharmaceutical feedstock production, the reactor will allow the study of irradiation in materials and the use of neutron beams to investigate the structures of materials in low dimensions [1, 2]. The RMB project has a suite of 15 instruments that use thermal and cold neutrons, allowing the investigation of structures at a wide range of micro- and nanoscales [2]. The Instituto de Pesquisas Energéticas e Nucleares (IPEN), in São Paulo - Brazil, houses the 5 MW research reactor IEA-R1. The correspondent reactor research center possesses a neutron radiography/tomography instrument that has made a great contribution to the Brazilian scientific community, since its creation in 1987 [3]. It has gone through recent upgrades that keep it in the select group of international instruments with a high thermal neutron flux at the sample position ($1.0 \times 10^{6}$ $n/cm^{2}s$) and, consequently, excellent spatial resolution [4, 5]. Even considering such excellence, we must consider that the IEA-R1 reactor is over 60 years old, which naturally makes RMB a successor facility that will bring many benefits to Brazilian society.

Like the other RMB instruments, the neutron imaging instrument (radiography and tomography) has been named after a bird of the Brazilian fauna. Neinei (scientific name = \textit{Megarynchus pitangua}) is a bird found in Brazil and belongs to the Tyrannidae family. Neinei will be located on the south face of RMB N01 building, where the proximity to the reactor pool (see Figure 1) will allow a better use of $1.0\times 10^{14}$ $n/cm^{2}s$ thermal neutron flux in the core. RMB has six times the power of IEA-R1, i.e, $30$ MW. Therefore, Neinei will enable research in hydrogenous substances [5], cultural heritage [4] and other materials that are analyzed using the traditional X-ray imaging technique [6].

Neutrons are spin $1/2$ neutral fermions that have great penetration into the matter because of their slight interaction with most chemical elements. However, neutrons interact strongly with a small group of light elements and distinguish isotopes of the same element. The use of polarized beams and spin projection, which is governed by the magnetic field of the nuclei that compose the sample, can reveal details of the structure of magnetic materials  [7]. The neutron imaging technique uses neutronic properties to study materials in $2D$ (radiography), and in $3D$ (tomography). Neutron imaging is a non-destructive technique that has been used to investigate metallic materials, ceramics, rocks and industrial material structures [4, 6, 7]. A disadvantage that neutron beam techniques posses in comparison to the X-ray technique is that nuclear facilities have high cost of design and construction. In this scenario, Monte Carlo simulations through the MCNP [8] and McStas [9] codes are essential to allow the development of instruments and optical components designs [10, 11, 12, 13].

The absence of other instruments nearby the Neinei Neutron imaging, according to the design stage of the RMB, brings the advantage of free assembly without spatial constraints, relying only on the physical limitation of the N01 building. The proximity to the reactor face allows the use of short lengths of neutron guides (for collimation purposes only) or tubes. According to Schillinger \textit{et al.} [14], neutron guides alone are not a good choice for a radiography setup because of the high divergence at the end of the guide ($L/D$ = 70 to 115, where $L$ is the distance between the slit, of diameter $D$, and the detector). According to the authors, the only way to obtain a parallel neutron beam is to use a slit and a long flight tube. If the beam area is to be covered, the arrangement can be followed by a neutron guide that carries the reactor flux within the acceptance angle of the guide. Recently, in the DINGO imaging instrument design project for the OPAL reactor, Garbe \textit{et al.} used an in-pile collimator to obtain a homogeneous neutron beam for two different configurations, at the end of a guide system (TG1) and beam port position (HB-1/2) for $L/D$ = $250$ and $1000$ [12, 13]. The design chosen for implementation in the project was HB-$1/2$, due to the flux of $4.75 \times 10^{7}$ $n/cm^{2}s$ in the detector position for $L/D$ = $250$, although the intensity distribution with the McStas software reveals inhomogeneous spatial distribution [12]. The uniformity of the beam incident on the sample directly affects the quality of the imaging technique [7, 11]. The unexpected inhomogeneous spatial distribution phenomenon, which always appears in the neutron radiography facilities installed at the end of the neutron guide, was investigated by Wang \textit{et al.} [11]. The authors report that the phenomenon observed is related to the spacing gaps in the guide transmission system and can be mitigated by decreasing the distances between the source and the guide entrance, increasing the guide length and source size, decreasing the guide cross-section [11]. The techniques for neutron radiography and tomography have improved in the last decades and the development of new instruments can be guided by all the knowledge presented in the literature [7, 15, 16, 17, 18].

In this spirit, the proposal of this work is the conception of the basic design of the Neinei instrument in RMB. Inspired by the works of Garbe \textit{et al.} [12], Wang \textit{et al.} [12] and Schillinger \textit{et al.} [14], we propose a basic configuration for Neinei using the McStas software [9]. In this proposal, we have investigated the uniformity and flux of the neutron beam at the detector position for $L/D$ ratios $= 100$, $500$ and $1000$. The results are compared to DINGO at OPAL [12, 13], Neutra at PSI [19], ANTARES at FRM-II [14] and BT2 at NIST [20]. Our results will have a direct impact on the Neinei instrument design project at more advanced stages.

\begin{figure}[hbt!]
\centering
\includegraphics[width=0.75\textwidth]{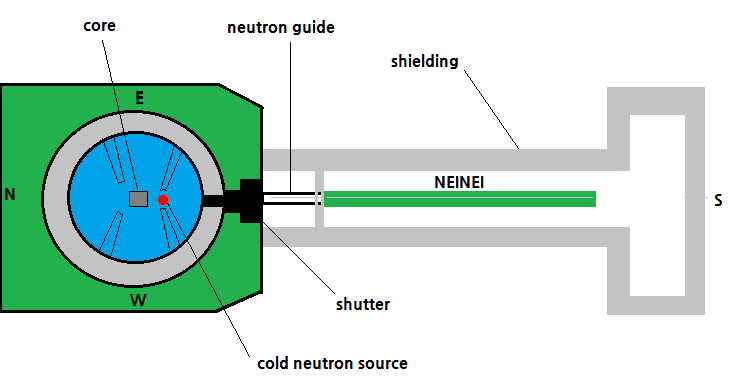}
\caption{Side-view sketch of the RMB and Neinei instrument. The remaining shutters and the neutron beam hall are on the west face of the reactor and are not shown to not overload the figure.}
\label{S_guide}
\end{figure}

\section{Monte Carlo Simulations of the RMB core}

The RMB reflector tank was modeled using the MCNP6 code, and a section along the Z-axis is shown in Figure 2. Highlighted in this scheme are the entrances of the thermal (TG) and cold (CG) neutron guides, the core reactor, the cold neutron source (CNS) and the neutron imaging entrance. The reactor core is composed of a 5x5 assembly of $U_{3}Si_{2}$ (19.75 wt percent) fuel elements, made up of 21 parallel plates framed by Aluminum $6061$ alloy. The simulations were carried out in the Nuclear Engineering Center (CEENG-IPEN), which has 200 processing nodes. Information about the distribution of neutrons such as position, velocity, spin vectors and their respective statistical weights are obtained from of PTRAC card [8].

\begin{figure}[hbt!]
\centering
\includegraphics[width=0.35\textwidth]{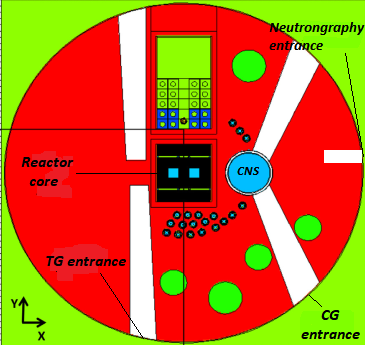}
\caption{Z-axis section of the RMB reflector tank modeled by MCNP6 code.}
\label{S_guide}
\end{figure}
The PTRAC file, which is an output from MCNP6, is used as an input file of the software McStas (Monte Carlo Simulation of Triple Axis Spectrometers) [9], in which we read the neutron beam through a virtual detector. We modeled the detector with an area of 100 $cm^{2}$ (10 cm x 10 cm) and a resolution of 2000 bins in order to obtain the intensity distribution found in Figure 3. There, the black dots represent the MCNP6 results, while the red curve is a Maxwellian fit, $\phi(\lambda )=(\eta/\lambda^5 T^2)exp(-\gamma/T\lambda^2)$, with parameters $\eta =(3.84056\pm0.00834)\times 10^{15}$ $n/cm^{2}sK^{2}\mathring{A}^{5}$, $T = (286.70239\pm0.4479)$ $K$, $\gamma = 933 \mathring{A}^{2}K$ and $R^{2} = 0.99772$. The Maxwellian $\phi(\lambda )$ is a function that describes the neutron distribution after thermal equilibrium with a material medium at temperature T [8]. Obtaining $\phi(\lambda )$ is a fundamental step in the design of nuclear facilities, as it allows evaluating the efficiency of neutron transport. 

\begin{figure}[hbt!]
\centering
\includegraphics[width=0.59\textwidth]{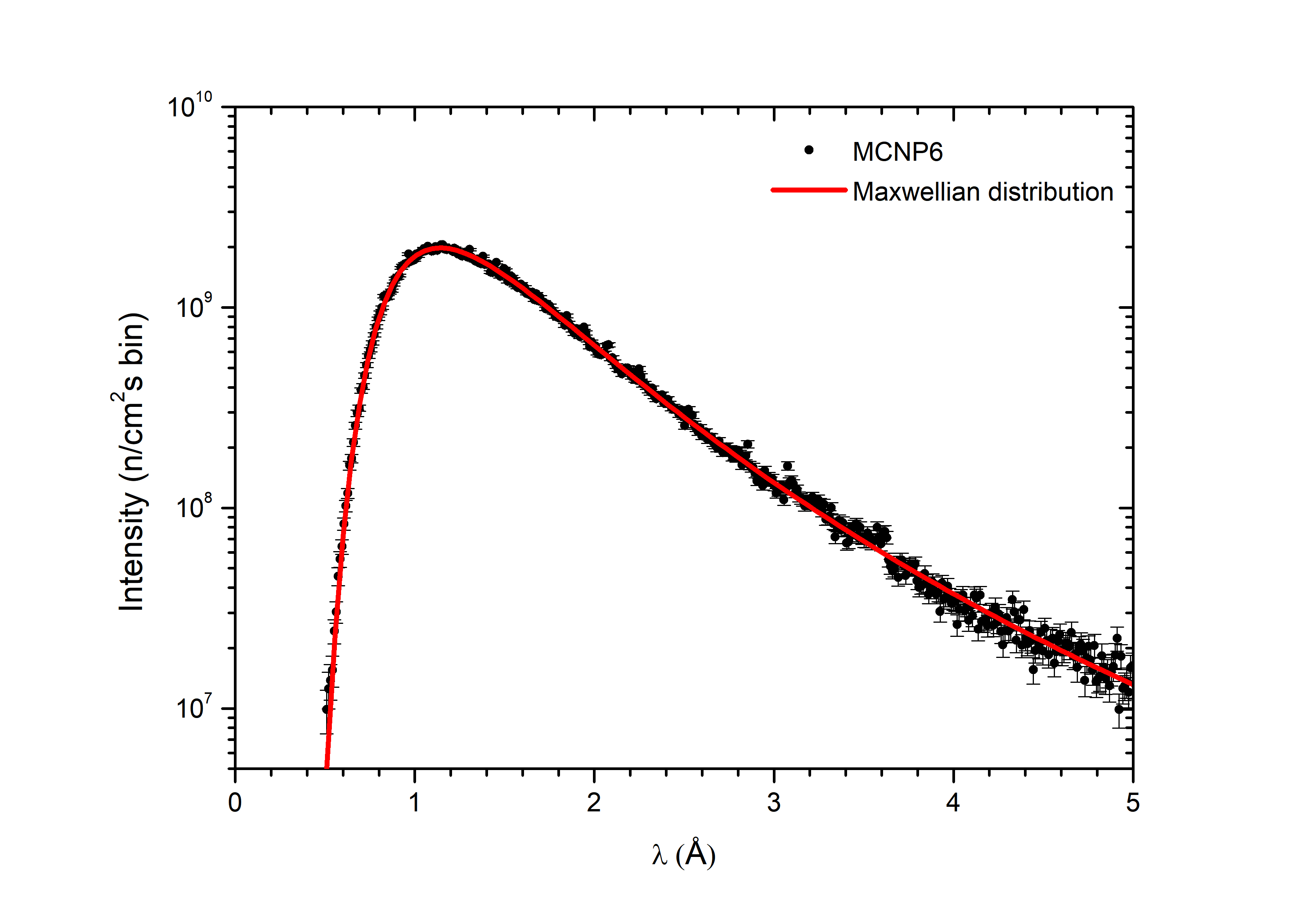}
\caption{Intensity of the neutron beam at the entrance to the neutrongraphy capture tube.}
\label{S_guide}
\end{figure}

\newpage

\section{Basic Neinei Design and McStas Simulations}

The basic design of the Neinei instrument is shown in Figure 4. Inspired by the DINGO instrument at OPAL, the neutron source for the instrument features an in-pile collimator in the primary shutter to homogenize the high-flux beam at the reactor face. To circumvent the inhomogeneous spatial distribution phenomenon, we have used a short neutron guide length according to the work of Wang \textit{et al.} [11] and Schillinger \textit{et al.} [14]. In Monte Carlo simulations, two values for guide length ($l_{g}$ $= 0.5$ m and $1.0$ m) and section of $50$ mm x $300$ mm are used to evaluate the beam divergence before it enters the instrument slit. In another configuration of guides for neutron transport, the use of a collimation length (extra guide) proved to be fruitful [10]. A slit selector allows the adoption of 3 different diameters, $D =$ $5$ mm, $10$ mm and $50$ mm, which allow obtaining $L/D$ ratios $100$, $500$ and $1000$, for $L = 5$ m, which is the same value adopted in the imaging instrument of IEA-R1 [4]. A rotating table with a stepper motor is used to obtain different angles of the sample in front of the incident neutron beam. The sample rotation adjustment will be determined according to the resolution of the CCD camera and the image storage system. The neutrons transmitted by the sample sensitize the scintillator (e.g., Li/ZnS), which generates 2D images captured by the CCD camera [5]. To avoid direct exposure of the CCD chip to radiation, a plane mirror reflects the image to the camera, which is installed at 90 degrees to the incident beam, as shown in Figure 4.

Monte Carlo simulations are performed using McStas ray tracing software [9]. A file containing the neutron distribution at the primary shutter position is generated through the PTRAC function of the MCNP code [8] for the RMB core. The neutron distribution is used by McStas as a source for the instrument to be designed. For this purpose, we have optical components arranged along the Z axis in the following sequence: \verb|Source_gen()|, \verb|Collimator_radial()|, \verb|Guide_simple()|, \verb|Slit()|, \verb|Absorber()|, \verb|PSD_monitor()|, and \verb|Monitor_nD()|. All components and parameters used for the simulations are listed in Table 1.

The simulations have been performed, and the results correspond to data collected in the detector at the end of the long flight box (absorber). Such data are the flux and the intensity of thermal neutrons, as a function of planar dimensions ($x$ and $y$), and are presented and discussed in the next section.

\begin{figure}[hbt!]
\centering
\includegraphics[width=0.65\textwidth]{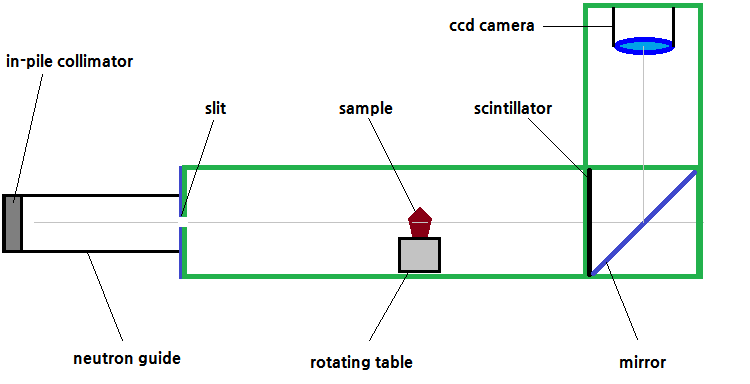}
\caption{Basic Neinei Design.}
\label{basic}
\end{figure}

\begin{table}[ht]
\centering
\caption{Simulation parameters of Neinei instrument.}
\begin{tabular}{cc}
\hline
\textbf{Components}  & \textbf{Parameters}                                                                                                                                                                       \\ \hline
Source        & \begin{tabular}[c]{@{}c@{}}RMB thermal beam, Intensity = $2.3 \times 10^{18}$; nps = $79909340$;\end{tabular}                           \\ \hline
Collimator         & \begin{tabular}[c]{@{}c@{}}Section (width x height) = 5 cm x 30 cm;\\ Length = 0.3 m; Divergence = 30';\end{tabular} \\ \hline
Neutron guide           & \begin{tabular}[c]{@{}c@{}}Section (width x height) = 5 cm x 30 cm;\\ Length = 0.5 m and 1.0 m;\\ Curvature = infinity;\\ Supermirror indexes: $M_{out} = 3$, $M_{top} = 3$, $M_{bottom} = 3$, $M_{in} = 3$;\end{tabular}                           \\ \hline
Slit   & \begin{tabular}[c]{@{}c@{}}Diameters = 5 mm, 10 mm and 50 mm;\end{tabular}                                                                                          \\ \hline
Absorbent box   & \begin{tabular}[c]{@{}c@{}}Section (width x height) = 20 cm x 30 cm;\\ Length = 5.0 m;\end{tabular}                                                                                          \\ \hline
Detector        & \begin{tabular}[c]{@{}c@{}}Section (width x height) = 5 cm x 30 cm; \end{tabular}          \\ \hline
\label{tab}
\end{tabular}
\end{table}

\section{Results}
\label{R}

The neutron imaging technique is sensitive to several parameters, including the $L/D$ resolution. A simple derivation allows us to obtain the beam divergence, given by $d = l/(L/D)$, where $l$ is the distance between the sample and the detector. Therefore, for the given detector resolution $C$, the other parameters must be adjusted by the constraint $C > l/(L/D)$ [17]. 

In Figures 5-10, we present the beam intensities as a function of the spatial coordinates for $l_{g} = 0.5$ m, $1.0$ m, and $L/D =$ $100$, $500$, and $1000$. We can see that the results reveal the effects of collimation by verifying a high neutron signal at the center of the detector. We found that the results for the guide length of $1.0$ m (Figures 8-10) provide the best neutron flux values in the detector. Therefore, we selected the results for $l_{g} =$ $1.0$ m for a comparative analysis presented in Table 2. We can observe that the results have excellent agreement with the instruments of other facilities if we compare the usual values for $L/D =$ $500$ and $1000$. Additionally, for testing purposes, we provided the value corresponding to the $L/D$ ratio = 100 only as an upper limit, finding a promising result that was physically expected. We have observed that even with the use of an in-pile collimator and an auxiliary neutron guide, before the slit, the neutron beam still presents inhomogeneity along the detection plane, as was observed by Garbe \textit{et al.} [12]. As pointed out by Wang \textit{et al.}, we observed a better distribution of the beam in the $L/D =$ 100 ratio in view of a larger slit opening and, consequently, better neutron sampling. If we compare it to the DINGO instrument from OPAL, the reference reactor for the RMB, we observe a difference of only $4$ percent in neutron flux for the high-resolution configuration. We also verified that all the results of our simulations show an expected decrease in intensity with an increase in the resolution of the instrument.
The neutron imaging technique is sensitive to several parameters, including the $L/D$ resolution. A simple derivation allows us to obtain the beam divergence, given by $d = l/(L/D)$, where $l$ is the distance between the sample and the detector. Therefore, for the given detector resolution $C$, the other parameters must be adjusted by the constraint $C > l/(L/D)$ [17]. 

A set of simulations reveals the behavior of the neutron flux in the instrument detector as a function of resolution (Figure 11). We can also observe that there is a decrease in the neutron flux of two orders of magnitude, which can be adjusted by the curve $\phi(L/D) = 10^{12.48921} \times (L/D)^{-1.81594}$. It is observed that the values obtained are found in intervals in which the parameters of the imaging instruments of the other installations are located, including the DINGO-OPAL.
As an initial study, the simulations have proven to provide guidelines for future, more in-depth studies where we will improve the instrument design.

\begin{table}[h]
\caption{Calculed flux compared with measured flux of neutron imaging instruments at OPAL, PSI, FRM2 and NIST $(n/cm^{2}s)$.}\label{table_L_P_L_S}
\begin{tabular*}{\tblwidth}{@{} LCCLCC@{} }
\toprule
\multirow{1}{*}{$L/D$} & Neinei (RMB) & DINGO (OPAL) & Neutra (PSI) & ANTARES (FRM2) & BT2 (NIST)  \\
\\
\midrule 

$100$ & $7.20 \times 10^{8}$ & $-$ & $-$ & $-$ & $-$ \\
$500$ & $3.64 \times 10^{7}$ & $4.75 \times 10^{7} (L/D=250)$ & $3.96 \times 10^{6} (L/D=550)$ & $9.4 \times 10^{7} (L/D=400)$ & $5.9 \times 10^{7} (L/D=400)$ \\
$1000$ & $1.10 \times 10^{7}$ & $1.15 \times 10^{7}$ & $NA$ & $2.5 \times 10^{7} (L/D=800)$ & $1.0 \times 10^{6} (L/D=1200)$ \\

\bottomrule
\end{tabular*}

\end{table}

\begin{figure}
     \centering
     \begin{subfigure}[t]{0.7\textwidth}
         \centering
         \includegraphics[width=\textwidth]{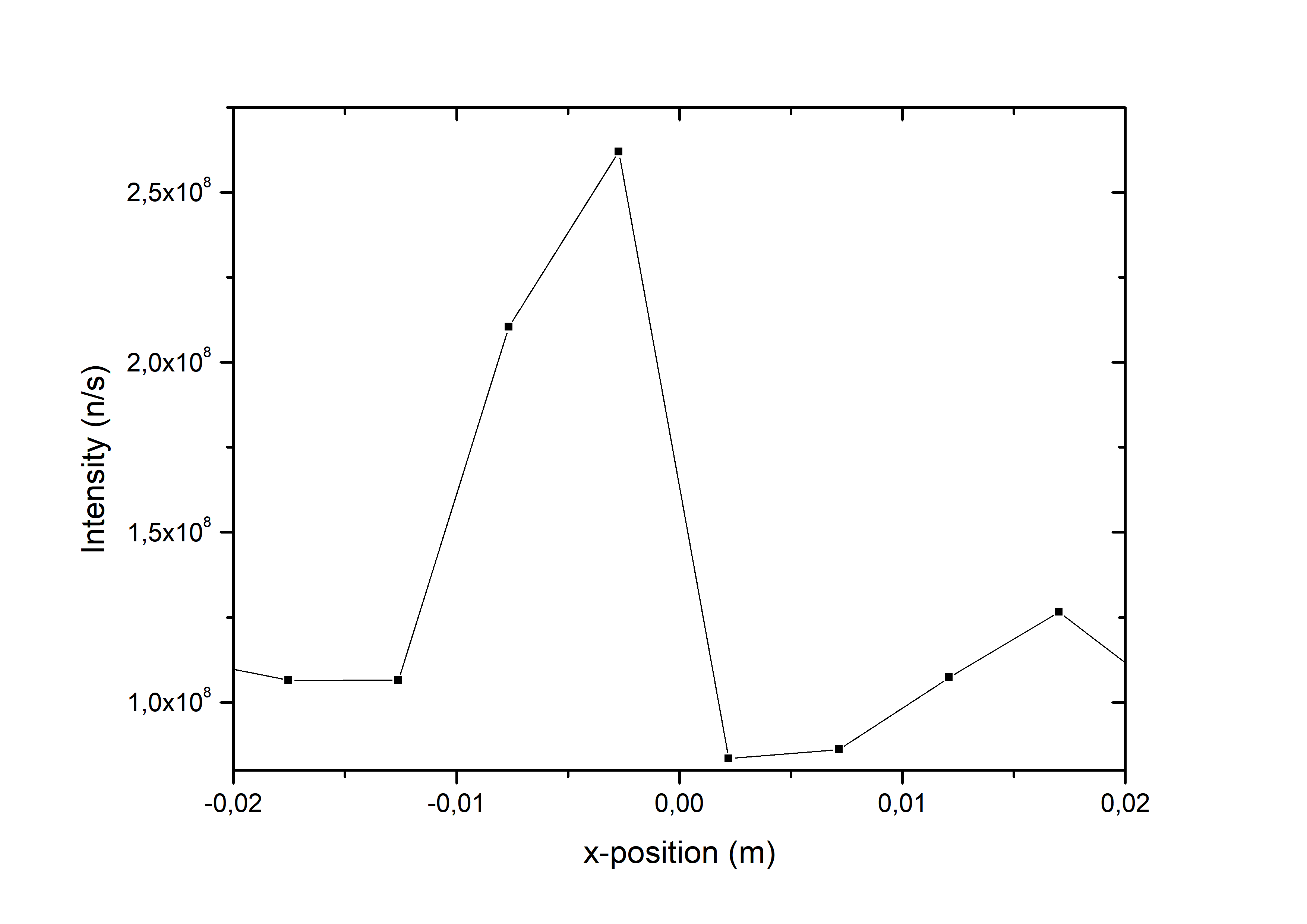}
         \label{um}
     \end{subfigure}
     \begin{subfigure}[b]{0.7\textwidth}
         \centering
         \includegraphics[width=\textwidth]{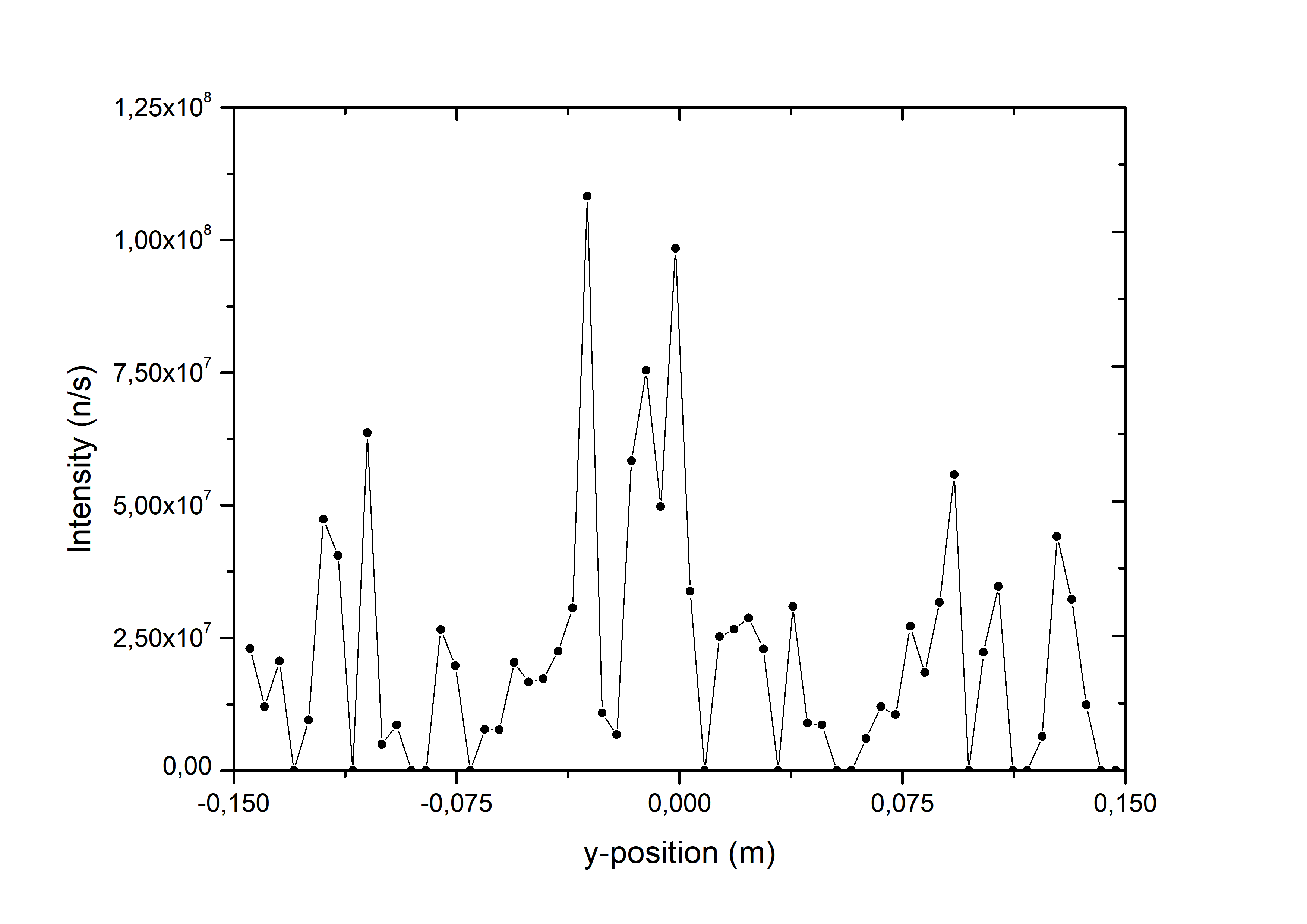}
         \label{dois}
     \end{subfigure}
        \caption{Calculated intensity distribution for $l_{g} = 0.5$ m and $L/D = 1000$ at the location of the detector, x-position (above) and y-position (below).}
        \label{fig:alfa}
\end{figure}

\begin{figure}
     \centering
     \begin{subfigure}[t]{0.7\textwidth}
         \centering
         \includegraphics[width=\textwidth]{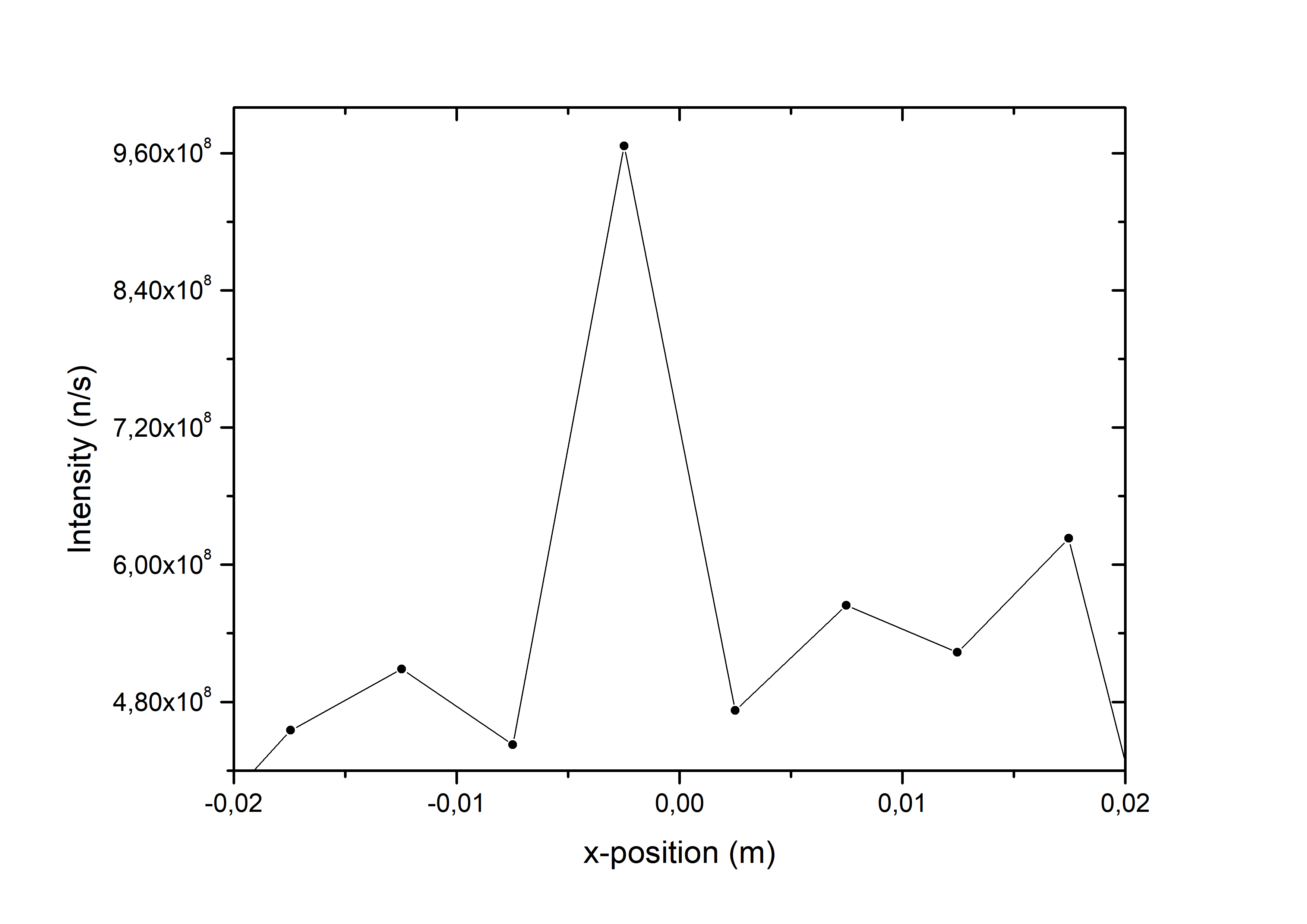}
         \label{tres}
     \end{subfigure}
     \begin{subfigure}[b]{0.7\textwidth}
         \centering
         \includegraphics[width=\textwidth]{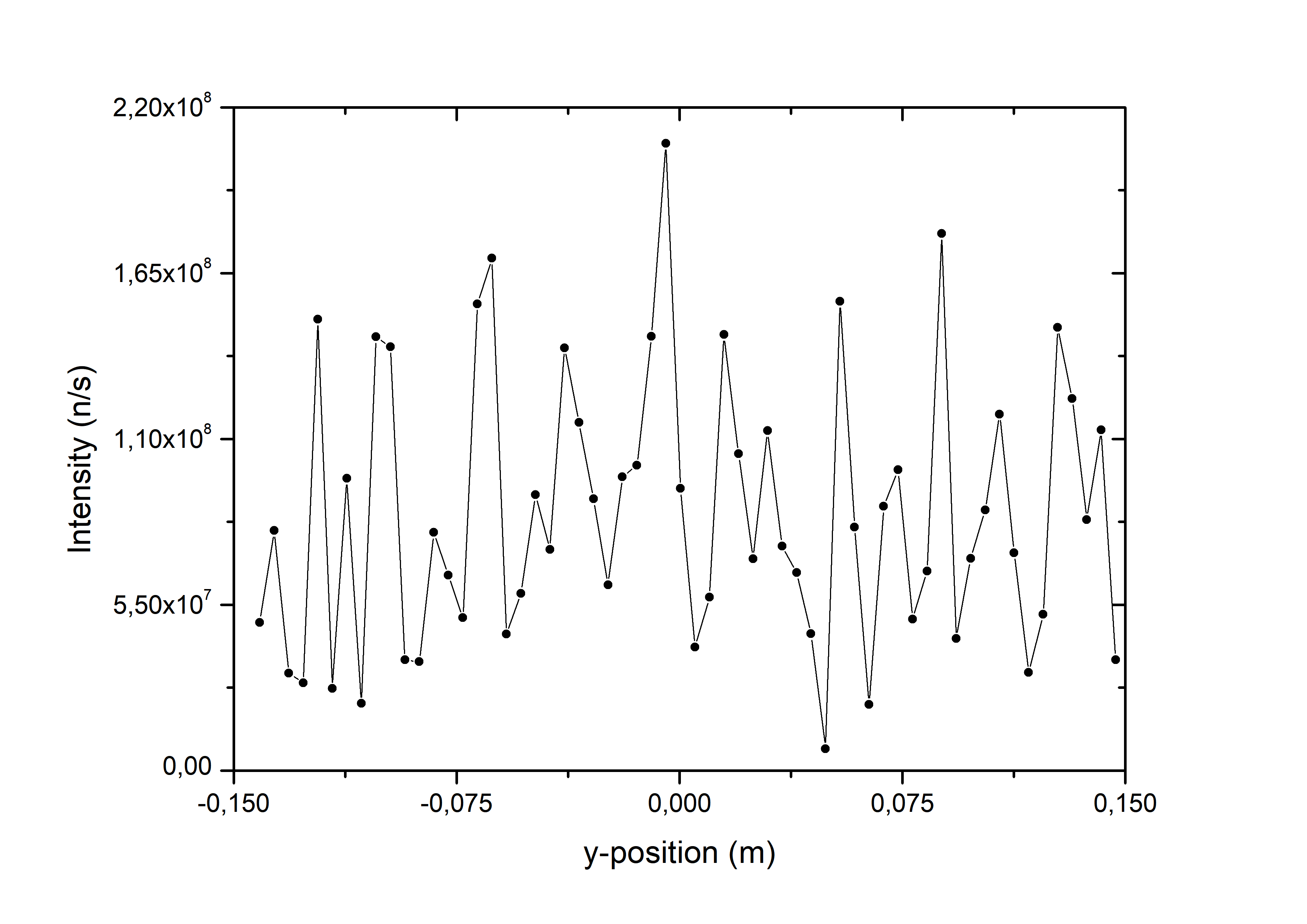}
         \label{quatro}
     \end{subfigure}
        \caption{Calculated intensity distribution for $l_{g} = 0.5$ m and $L/D = 500$ at the location of the detector, x-position (above) and y-position (below).}
        \label{fig:beta}
\end{figure}

\begin{figure}
     \centering
     \begin{subfigure}[t]{0.7\textwidth}
         \centering
         \includegraphics[width=\textwidth]{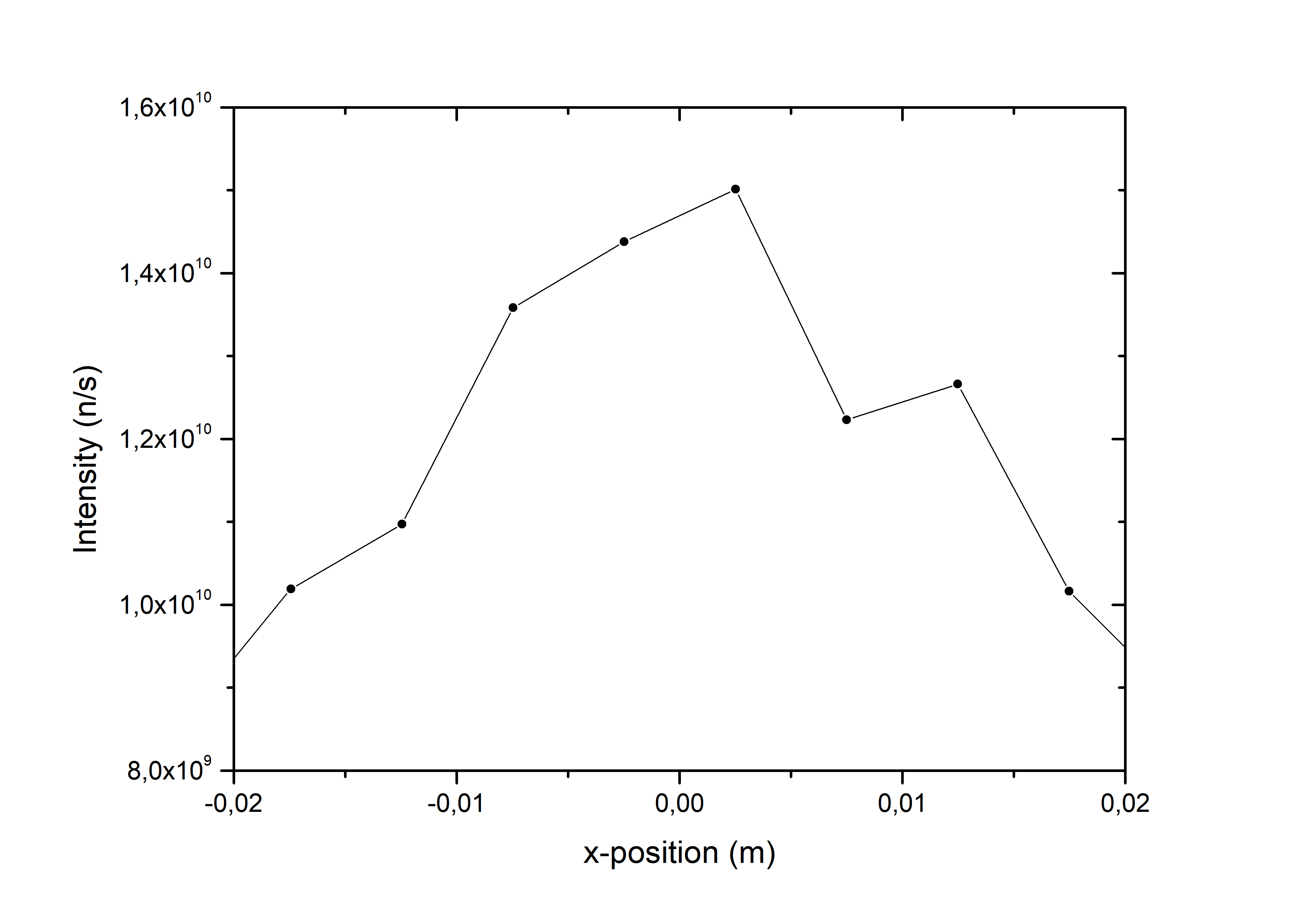}
         \label{cinco}
     \end{subfigure}
     \begin{subfigure}[b]{0.7\textwidth}
         \centering
         \includegraphics[width=\textwidth]{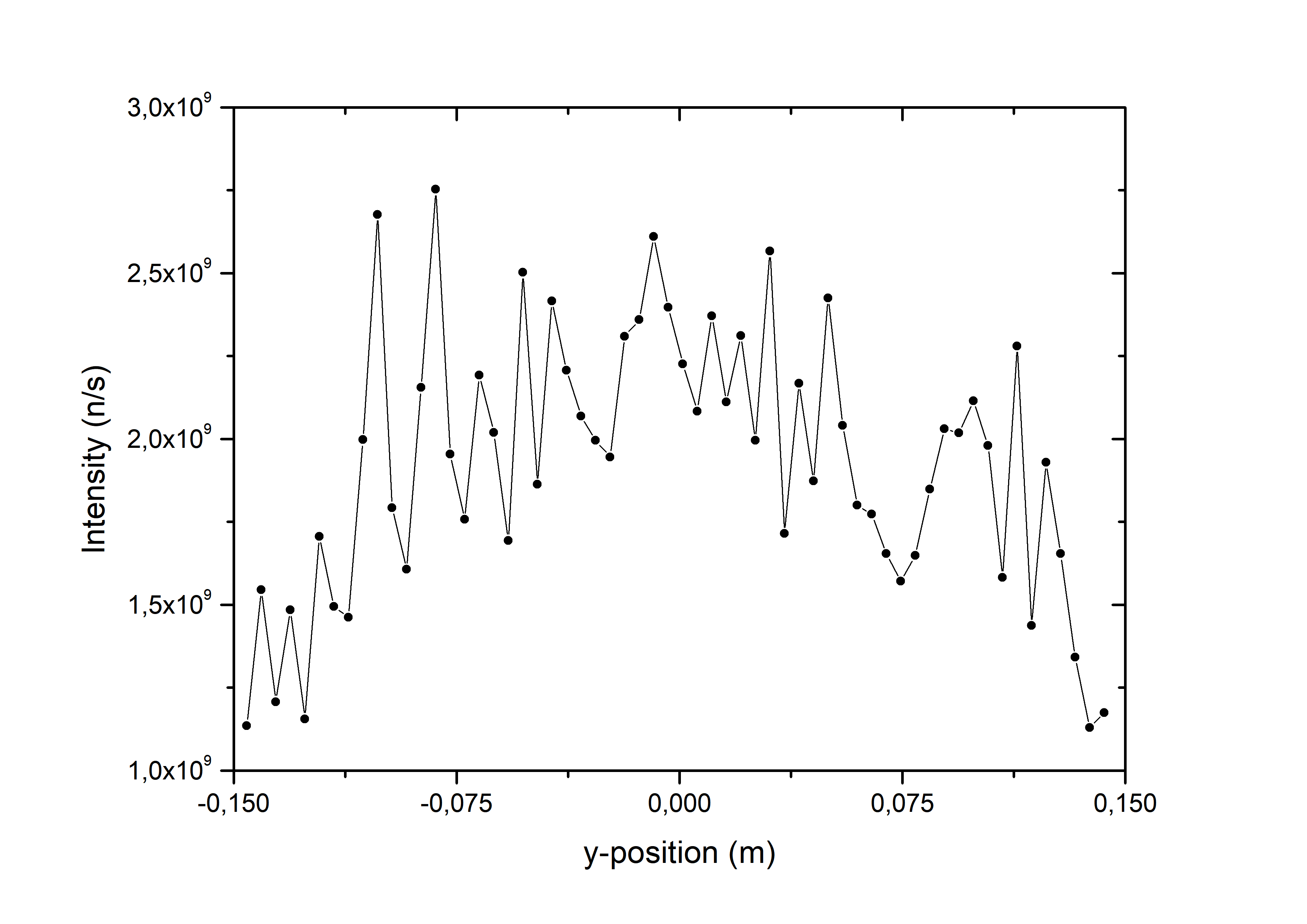}
         \label{seis}
     \end{subfigure}
        \caption{Calculated intensity distribution for $l_{g} = 0.5$ m and $L/D = 100$ at the location of the detector, x-position (above) and y-position (below).}
        \label{fig:gamma}
\end{figure}

\begin{figure}
     \centering
     \begin{subfigure}[t]{0.7\textwidth}
         \centering
         \includegraphics[width=\textwidth]{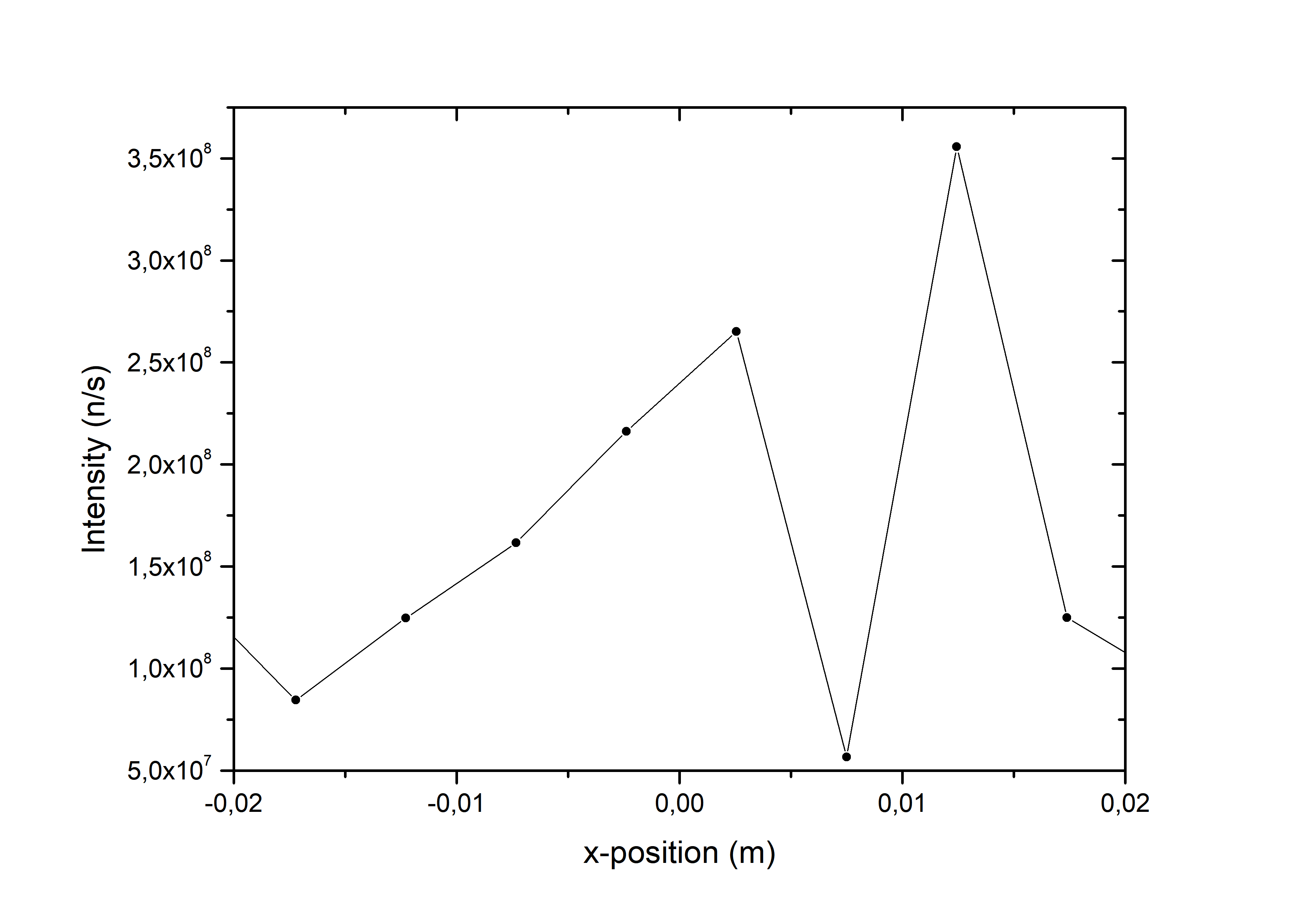}
         \label{sete}
     \end{subfigure}
     \begin{subfigure}[b]{0.7\textwidth}
         \centering
         \includegraphics[width=\textwidth]{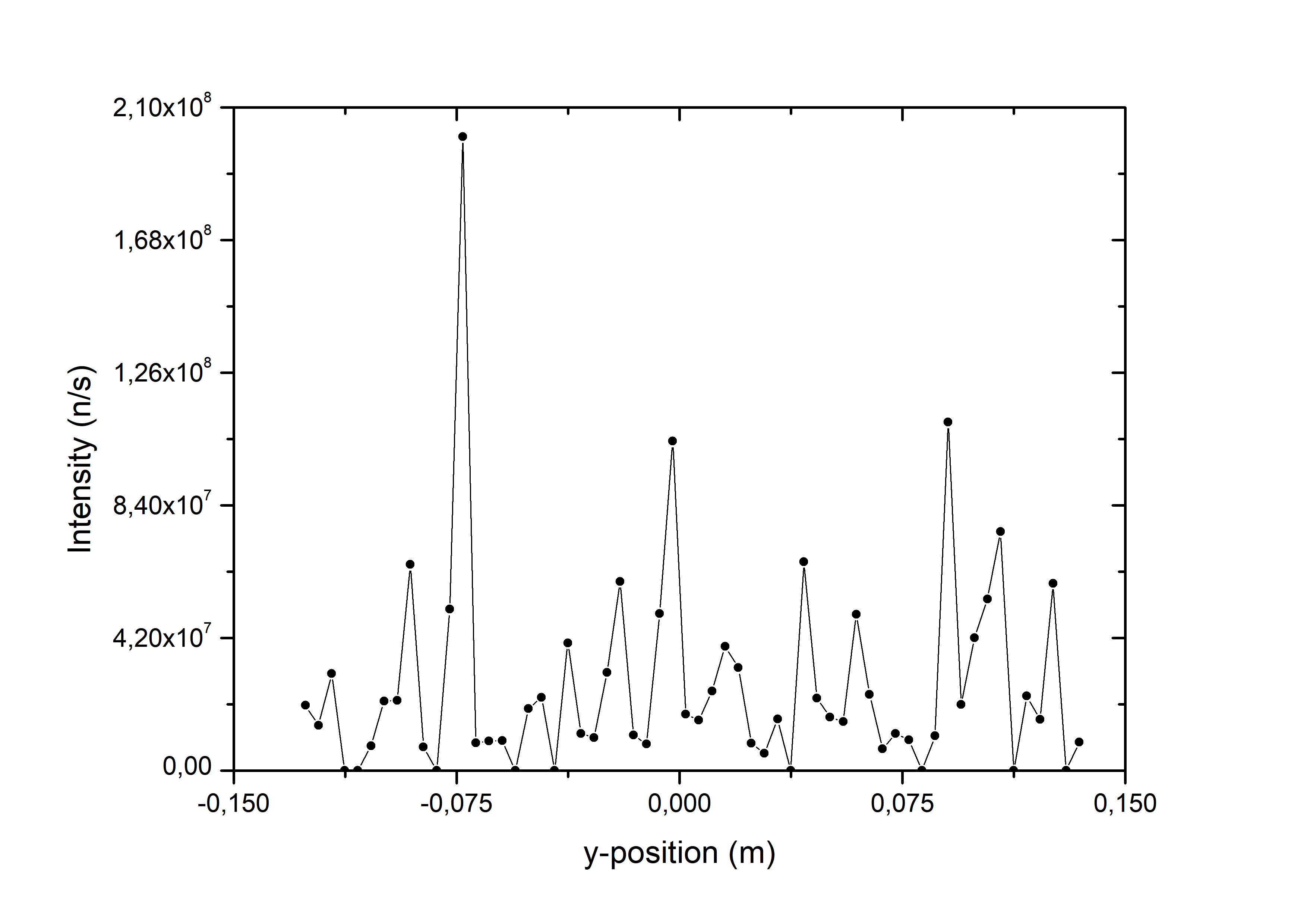}
         \label{oito}
     \end{subfigure}
        \caption{Calculated intensity distribution for $l_{g} = 1.0$ m and $L/D = 1000$ at the location of the detector, x-position (above) and y-position (below).}
        \label{fig:delta}
\end{figure}

\begin{figure}
     \centering
     \begin{subfigure}[t]{0.7\textwidth}
         \centering
         \includegraphics[width=\textwidth]{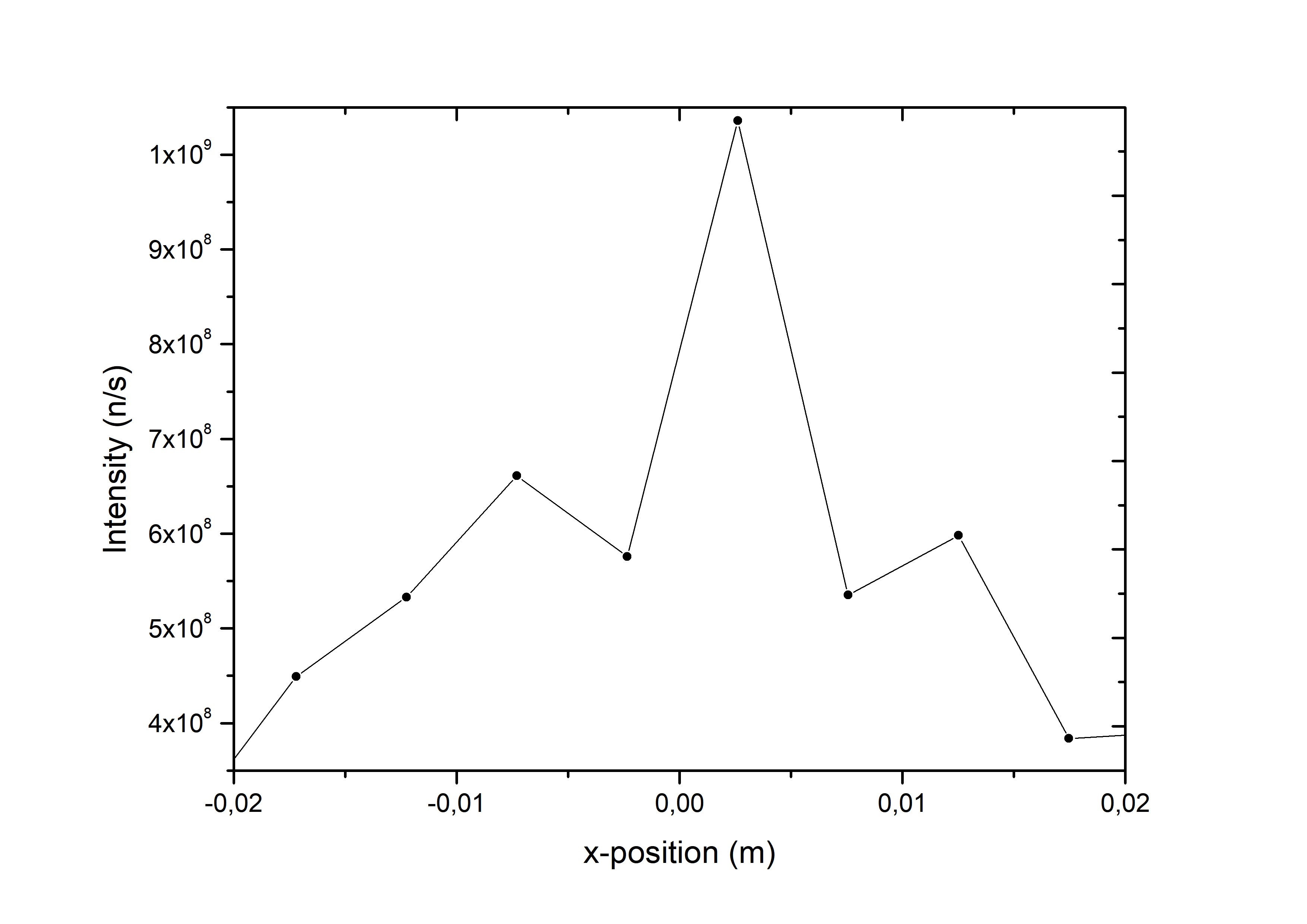}
         \label{nove}
     \end{subfigure}
     \begin{subfigure}[t]{0.7\textwidth}
         \centering
         \includegraphics[width=\textwidth]{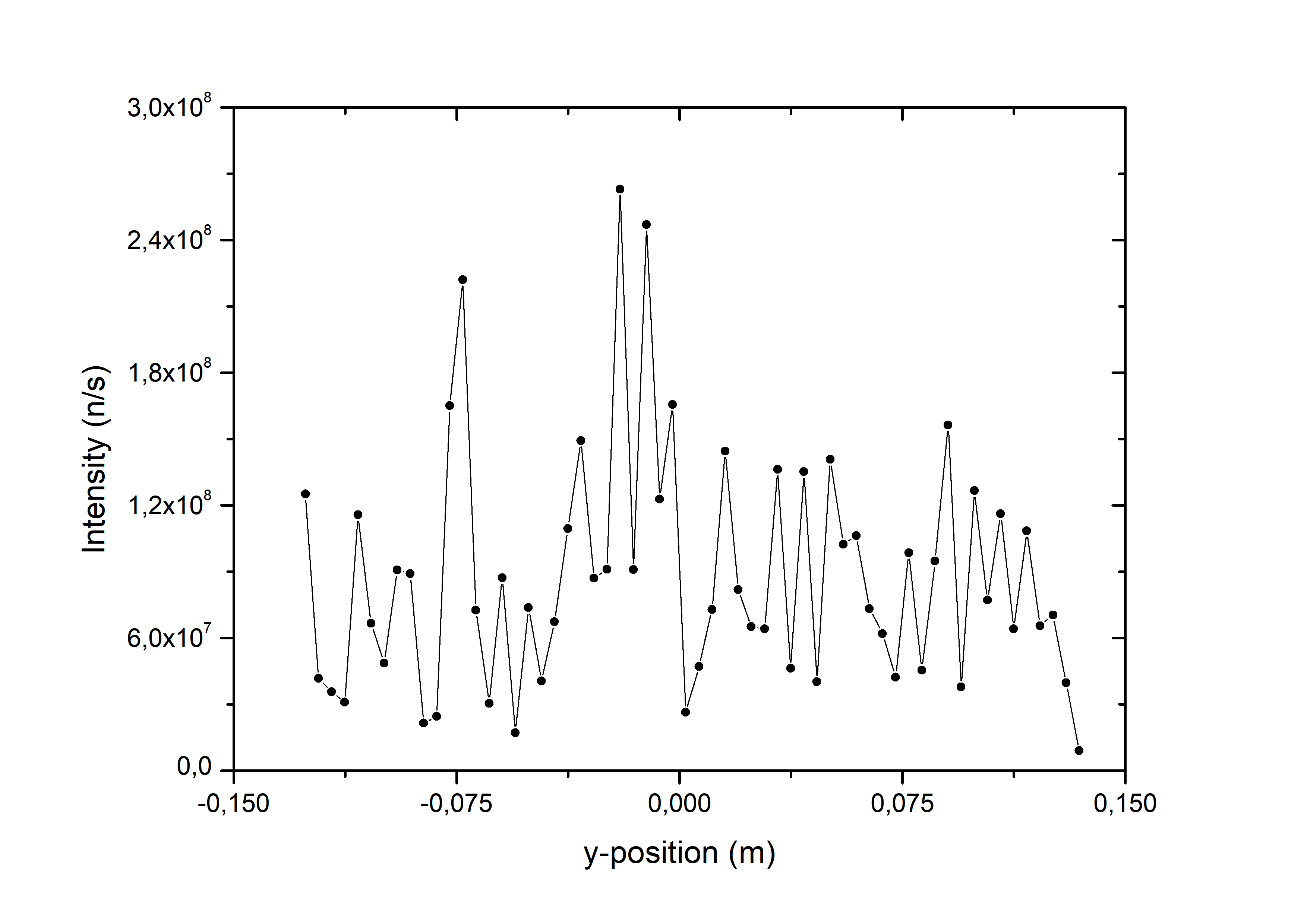}
         \label{dez}
     \end{subfigure}
        \caption{Calculated intensity distribution for $l_{g} = 1.0$ m and $L/D = 500$ at the location of the detector, x-position (above) and y-position (below).}
        \label{fig:zeta}
\end{figure}

\begin{figure}
     \centering
     \begin{subfigure}[t]{0.7\textwidth}
         \centering
         \includegraphics[width=\textwidth]{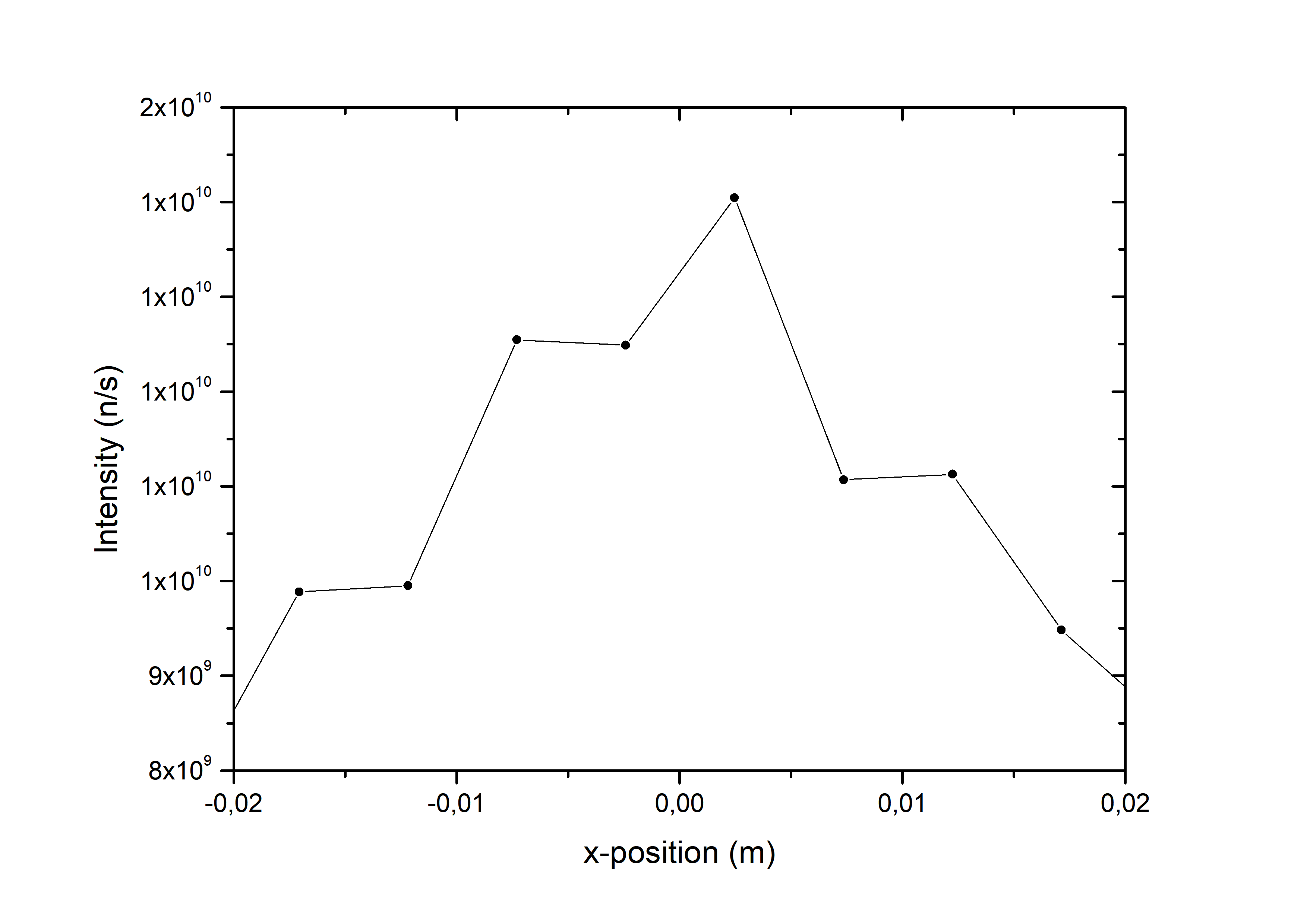}
         \label{onze}
     \end{subfigure}
     \begin{subfigure}[t]{0.7\textwidth}
         \centering
         \includegraphics[width=\textwidth]{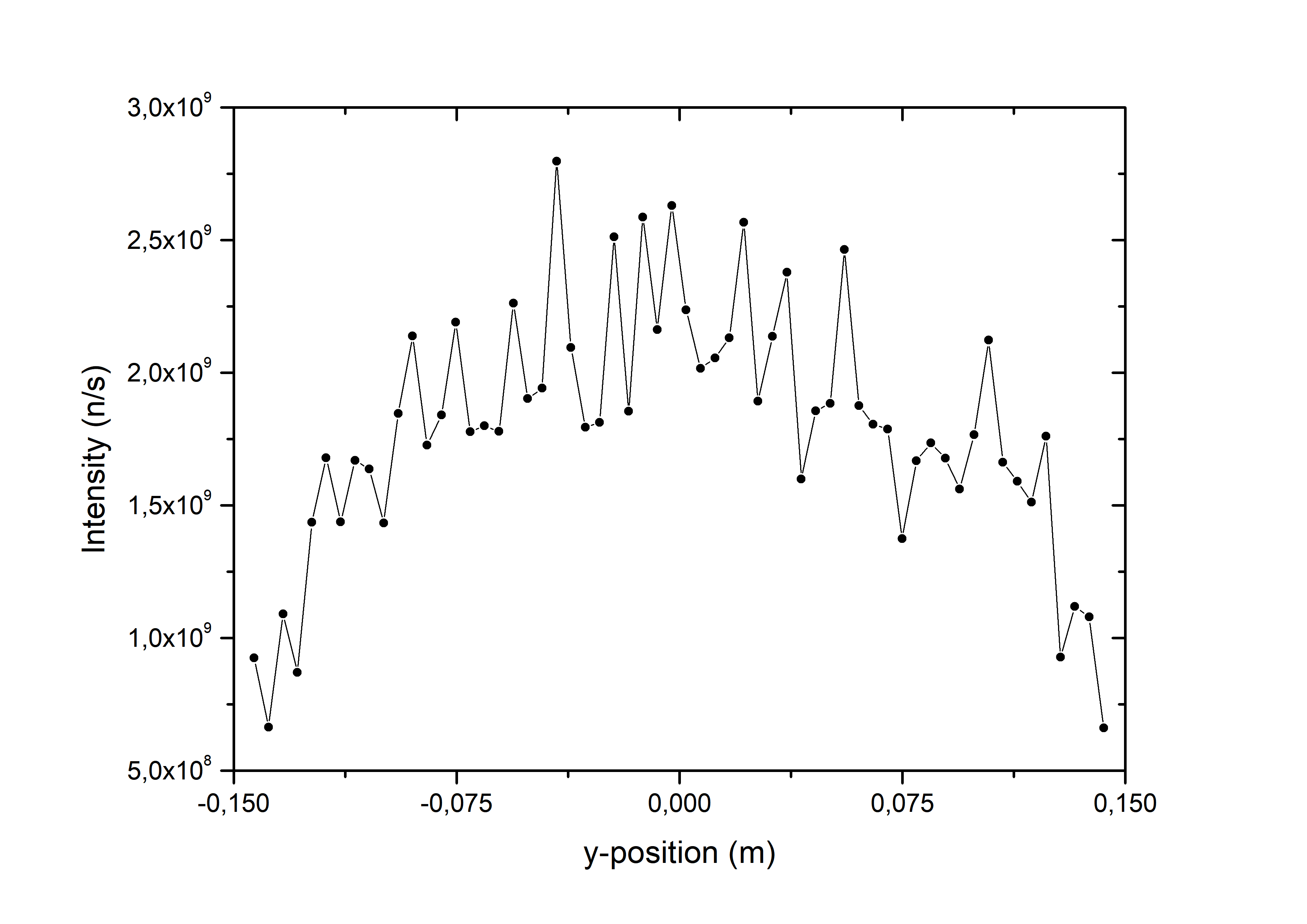}
         \label{doze}
     \end{subfigure}
        \caption{Calculated intensity distribution for $l_{g} = 1.0$ m and $L/D = 100$ at the location of the detector, x-position (above) and y-position (below).}
        \label{fig:eta}
\end{figure}

\begin{figure}[t]
\centering
\includegraphics[width=0.65\textwidth]{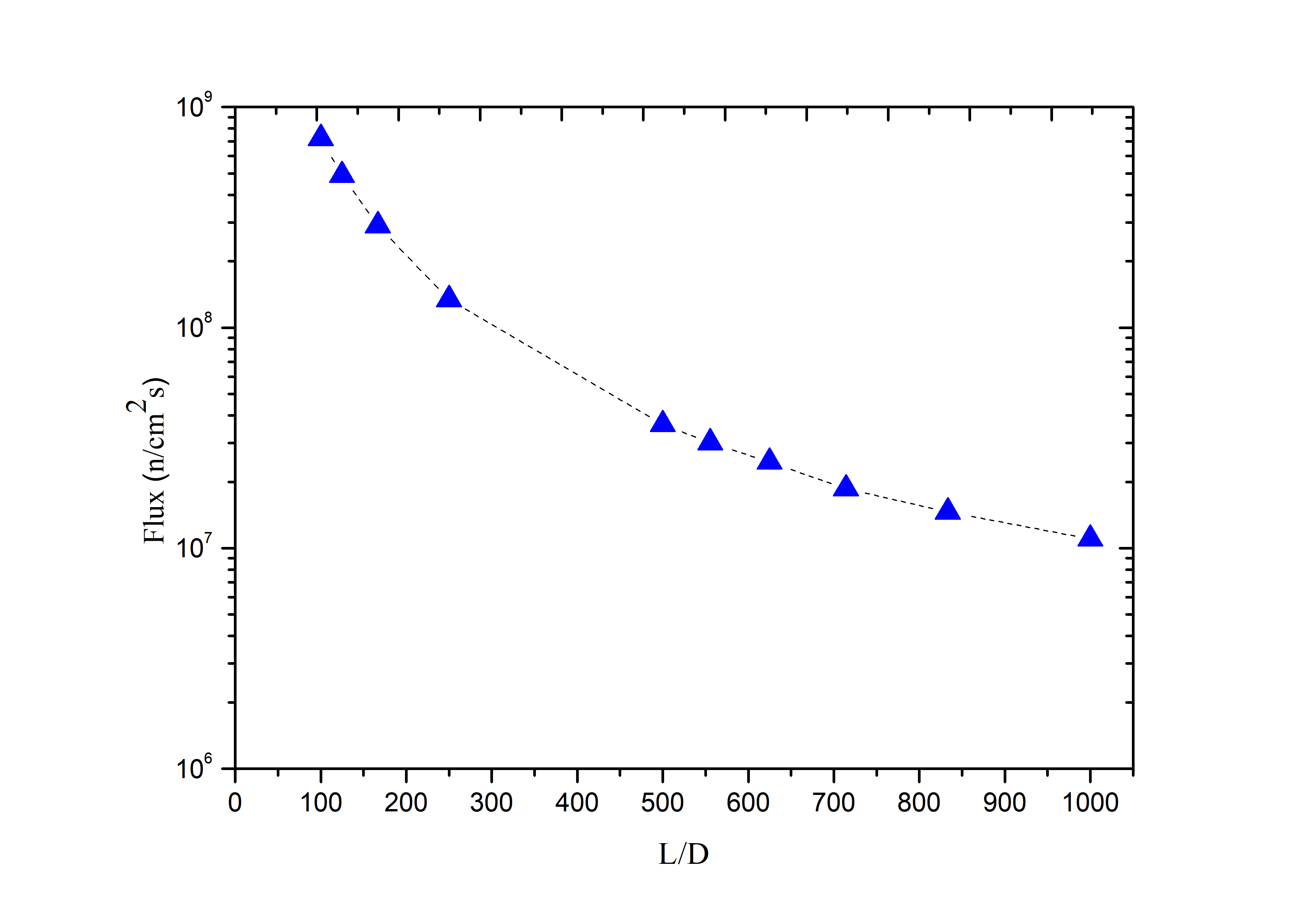}
\caption{Neutron flux in the detector as a function of Neinei resolution. The $L/D$ range covers all neutron imaging instruments compared.}
\label{complementar}
\end{figure}
\newpage

\newpage

\newpage

\newpage

\newpage

\newpage

\pagebreak[4]

\section{Conclusions}
\label{C}

The beam intensities found are in agreement with the results of other neutron imaging instruments [12, 14, 15, 20]. Additionally, we provide the results referring to a low-resolution configuration, which provides excellent signal intensity, as physically expected. It is worth mentioning that the presented model still shows an inhomogeneous neutron beam, even after a collimation system formed by an in-pile collimator and an auxiliary neutron guide. However, this phenomenon is found in the literature and is associated with spatial gaps between the components of the imaging instrument [11], although there are other geometric constraints that can cause beam divergence [14].
\newpage
In general, our results are promising and indicate important directions for the continuity of the project. In terms of equipment shielding, it must be pointed out that the McStas software cannot estimate the dose of gamma radiation, both in the region of the detectors and in the vicinity of the primary shutter. Therefore, a more detailed study with the MCNP code should be carried out for details of shielding the Neinei instrument. The sensitivity of the camera's CCD chip requires special care, and the use of filters (e.g., bismuth) in the beam path can be very important.

\newpage

\newpage

\newpage
\newpage
\newpage
\newpage
\pagebreak[4]
\newpage
\section*{Acknowledgement}
L.P. de Oliveira also would like to thank National Nuclear Energy Commission (CNEN) for financial support, Process SEI 01341011049/2021-22. 
\newpage

\newpage

\newpage

\newpage 
\newpage

\newpage
\newpage

\newpage
\newpage

\newpage

\pagebreak[2]

\newpage
\newpage
\section*{References}

\noindent\label{1} [1] J. Perrotta and A. Soares, \textit{RMB: the new brazilian multipurpose research reactor}, \textit{International Journal for Nuclear Power} 60 (2015) 30.\\
\noindent [2] A. Souza, L. de Oliveira, F. Yokaichiya, F. Genezini and M. Franco, \textit{Neutron guide building instruments of the Brazilian Multipurpose Reactor (RMB) project}, \textit{Journal of Instrumentation} 15 (2020) P04011.\\
\noindent [3] F. Souza, G. Zahn and P. Silva, \textit{Contribuições do Reator IEA-R1 para a Pesquisa Nuclear: Workshop Anual do Reator de Pesquisas – WARP 2}, vol. 1. Editora Blucher, 2022, \textit{http://dx.doi.org/10.5151/9786555501483}.\\
\noindent [4] M. S. Pereira, R. Schoueri, C. Domienikan, F. de Toledo, M. Andrade and R. Pugliesi, \textit{The neutron tomography facility of ipen-cnen/sp and its potential to investigate ceramic objects from the brazilian cultural heritage}, \textit{Applied Radiation and Isotopes} 75 (2013) 6.\\
\noindent [5] R. Schoueri, C. Domienikan, F. de Toledo, M. Andrade, M. Stanojev Pereira and R. Pugliesi, \textit{The new facility for neutron tomography of ipen-cnen/sp and its potential to investigate hydrogenous substances}, \textit{Applied Radiation and Isotopes} 84 (2014) 22.\\
\noindent [6] J. Banhart, A. Borbély, K. Dzieciol, F. Garcia-Moreno, I. Manke, N. Kardjilov et al., \textit{X-ray and neutron imaging – complementary techniques for materials science and engineering: Dedicated to professor dr. h.-p. degischer on the occasion of his 65th birthday}, \textit{International Journal of Materials Research} 101 (2010) 1069.\\
\noindent [7] N. Kardjilov, I. Manke, A. Hilger, M. Strobl and J. Banhart, \textit{Neutron imaging in materials science}, \textit{Materials Today} 14 (2011) 248.\\
\noindent [8] T. Goorley, M. James, T. Booth, F. Brown, J. Bull, L. Cox et al., \textit{Features of MCNP6}, \textit{Annals of Nuclear Energy} 87 (2016) 772.\\
\noindent [9] K. Lefmann and K. Nielsen, \textit{MCSTAS, a general software package for neutron ray-tracing simulations}, \textit{Neutron News} 10 (1999) 20.\\
\noindent [10] A. Souza, L. de Oliveira and F. Genezini, \textit{Monte carlo simulations of the s-shaped neutron guides with asymmetric concave and convex surface coatings}, \textit{Nuclear Instruments and Methods in Physics Research Section A: Accelerators, Spectrometers, Detectors and Associated Equipment} 1031 (2022) 166607.\\
\noindent [11] Y. Wang, G. Wei, H. Wang, Y. Liu, L. He, K. Sun et al., \textit{A study on inhomogeneous neutron intensity distribution origin from neutron guide transportation}, \textit{Physics Procedia} 88 (2017) 354.\\
\noindent [12] U. Garbe, T. Randall and C. Hughes, \textit{The new neutron radiography/tomography/imaging station dingo at opal}, \textit{Nuclear Instruments and Methods in Physics Research Section A: Accelerators, Spectrometers, Detectors and Associated Equipment} 651 (2011) 42.\\
\noindent [13] U. Garbe, T. Randall, C. Hughes, G. Davidson, S. Pangelis and S. Kennedy, \textit{A new neutron radiography / tomography / imaging station dingo at opal}, \textit{Physics Procedia} 69 (2015) 27.\\
\noindent [14] B. Schillinger, E. Calzada, F. Grünauer and E. Steichele, \textit{The design of the neutron radiography and tomography facility at the new research reactor frm-ii at technical university munich}, \textit{Applied Radiation and Isotopes} 61 (2004) 653.\\
\noindent [15] A. Hilger, N. Kardjilov, M. Strobl, W. Treimer and J. Banhart, \textit{The new cold neutron radiography and tomography instrument conrad at hmi berlin}, \textit{Physica B: Condensed Matter} 385-386 (2006) 1213.\\
\noindent [16] I. Manke, N. Kardjilov, A. Hilger, M. Strobl, M. Dawson and J. Banhart, \textit{Polarized neutron imaging at the conrad instrument at helmholtz centre berlin}, \textit{Nuclear Instruments and Methods in Physics Research Section A: Accelerators, Spectrometers, Detectors and Associated Equipment} 605 (2009) 26.\\
\noindent [17] N. Kardjilov, E. Lehmann, E. Steichele and P. Vontobel, \textit{Phase-contrast radiography with a polychromatic neutron beam}, \textit{Nuclear Instruments and Methods in Physics Research Section A: Accelerators, Spectrometers, Detectors and Associated Equipment} 527 (2004) 519.\\
\noindent [18] N. Kardjilov, S. Baechler, M. Bastürk, M. Dierick, J. Jolie, E. Lehmann et al., \textit{New features in cold neutron radiography and tomography part ii: applied energy-selective neutron radiography and tomography}, \textit{Nuclear Instruments and Methods in Physics Research Section A: Accelerators, Spectrometers, Detectors and Associated Equipment} 501 (2003) 536.\\
\noindent [19] E. Lehmann, H. Pleinert and L. Wiezel, \textit{Design of a neutron radiography facility at the spallation source sinq}, \textit{Nuclear Instruments and Methods in Physics Research Section A: Accelerators, Spectrometers, Detectors and Associated Equipment} 377 (1996) 11.\\
\noindent [20] D. Hussey, D. Jacobson, M. Arif, P. Huffman, R. Williams and J. Cook, \textit{New neutron imaging facility at the nist}, \textit{Nuclear Instruments and Methods in Physics Research Section A: Accelerators, Spectrometers, Detectors and Associated Equipment} 542 (2005) 9.


@article{mcstas,
  doi = {10.1080/10448639908233684},
  url = {https://doi.org/10.1080/10448639908233684},
  year = {1999},
  month = jan,
  publisher = {Informa {UK} Limited},
  volume = {10},
  number = {3},
  pages = {20--23},
  author = {Kim Lefmann and Kristian Nielsen},
  title = "{{MCSTAS},  a general software package for neutron ray-tracing simulations}",
  journal = {Neutron News}
}
\end{document}